\documentclass[aps,twocolumn,groupedaddress,prb,floatfix,showpacs]{revtex4}
\usepackage{graphicx}
\usepackage{dcolumn}
\usepackage{bm}
\usepackage{amsmath}
\usepackage{verbatim}
\begin{document}

\title{Effects of edge magnetism and external electric field on energy gaps in multilayer graphene nanoribbons}

\author{Bhagawan Sahu$^1$}
\email{brsahu@physics.utexas.edu}
\author{Hongki Min$^{2,3}$}
\author{Sanjay K. Banerjee$^1$}
\affiliation{
$^1$Microelectronics Research Center, The University of Texas at Austin, Austin, Texas 78758, USA\\
$^2$Center for Nanoscale Science and Technology, National Institute of Standards and Technology, Gaithersburg, Maryland 20899-6202, USA\\
$^3$Maryland NanoCenter, University of Maryland, College Park, Maryland 20742, USA
}

\date{\today}

\begin{abstract}
Using 
\textit{first principles} 
density functional theory, we study the electronic structure of multilayer graphene nanoribbons as a function of 
the ribbon width and the external electric field, applied perpendicular to the ribbon layers. 
We consider two types of edges (armchair and zigzag), each with two edge alignments (referred to as $\alpha$- and $\beta$-alignments). 
We show that, as in monolayer and bilayer armchair nanoribbons, multilayer armchair nanoribbons exhibit three classes of energy gaps which decrease with increasing
width.
Non-magnetic multilayer zigzag nanoribbons have band structures that are sensitive to the edge alignments and the number of layers, indicating different magnetic properties and resulting energy gaps. We find that energy gaps can be induced in ABC-stacked ribbons with an perpendicular external electric field, while in other stacking sequences, the gaps decrease or remain closed as the external electric field increases.
\end{abstract}

\pacs{71.15.Mb, 81.05.Uw, 75.75.+a}
\maketitle

\section{Introduction}
Recent progress in the isolation of single and multilayer graphene sheets has opened up a new topic in two-dimensional electron systems \cite{geim2007a,geim2007b,castro_neto2009}. Electronic structures of single and multilayer graphene \cite{hongki1} continue to contribute 
interesting physics and hint at possible practical applications in carbon based electronics \cite{sanjay,hongki2,hongki3,hongki4}. Recent predictions in multilayer graphene address chiral decomposition of electronic states in multilayer graphene stacks \cite{hongki1}, energy-gap opening only in ABC-periodic graphene stacks with an perpendicular external electric field applied perpendicularly to the ribbon layers, and importance of electron-electron interactions in multilayer systems due to the appearance of flat bands near the Fermi level \cite{fan}. It is interesting to study whether some of these predictions can be extended to finite size multilayer graphene stacks (such as ribbons and flakes). 
Moreover,
due to recent advances in fabrication of less than 10 nm wide nanoribbons with the control of their edge morphology \cite{dai,cai,milli1} and observing magnetic edge states in few-layer graphene ribbons \cite{milli2,geim2010}, our studies of
interplay of magnetism and electric field effects in multilayer nanoribbons have important implications in interpreting experiments. 

In this paper, we report on the electronic structure of multilayer graphene nanoribbons using 
\textit{first principles}
electronic structure method \cite{paolo} and address the interplay of magnetism, 
perpendicular external electric field and the energy gap. 
Our study suggests the existence of three classes of energy gaps in multilayer armchair nanoribbons, and strong dependence of magnetic properties on the edge alignments and the number of layers in multilayer zigzag nanoribbons. We discuss the effect of a perpendicular external electric field on multilayer nanoribbons and find that gaps can be enhanced in metallic ABC-stacked nanoribbons whereas in other nanoribbons, gaps decrease with increasing electric field or remain almost zero.

We begin by describing the computational method and the parameters used for this study in section II. In section III, we present the electronic structure of multilayer armchair nanoribbons and discuss the width dependence of energy gaps. Band structure of multilayer zigzag nanoribbons and the interplay of band structure, edge magnetism and resulting gaps will be discussed in section IV. In section V, we discuss the external electric field effects on the energy gaps of AB- and ABC-periodic ribbons. In the following discussions, the direction of the applied external electric field is taken to be perpendicular to the ribbon layers.  Finally we present our summary and conclusions.

\begin{figure}[ht]
\includegraphics[width=1\linewidth]{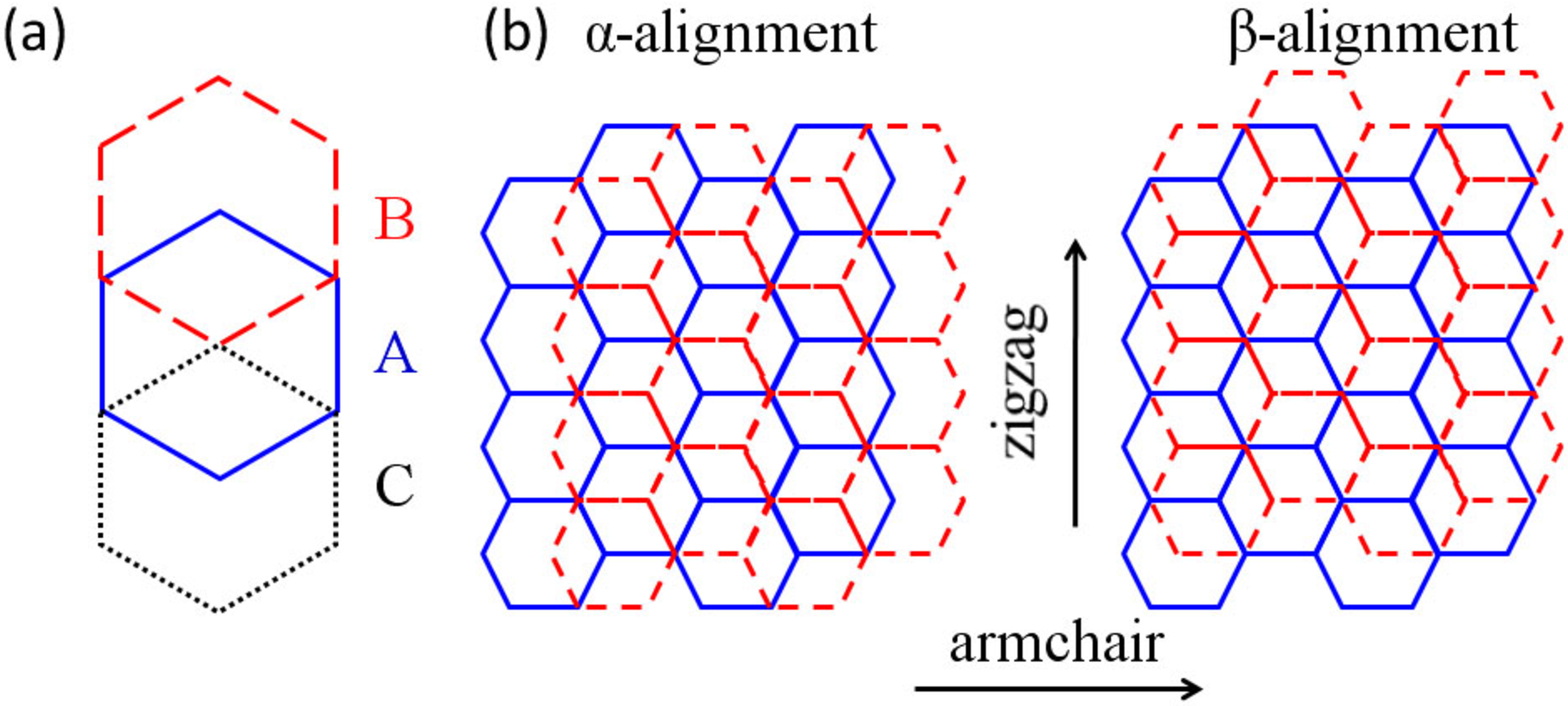}
\caption{ (Color online) Schematic illustration of (a) three types of stacking arrangements, labeled by A, B and C, and (b) two types of edge alignments, $\alpha$-alignment and $\beta$-alignment in multilayer graphene nanoribbons. 
The two edge alignments are distinguished by different ways of shifting the top layer with respect to the other in finite size multilayer graphene stacks.
The arrows indicate the direction of edges along which nanoribbons span infinitely.
}
\label{fig:fig1}
\end{figure}  

\section{Density Functional Theory Calculations} 

We use a plane-wave basis-based electronic structure method \cite{paolo} with ultrasoft pseudopotentials \cite{ultrasoft} for core-valence interaction to obtain multilayer band structures. In our previous studies of bilayer nanoribbons \cite{sahu1} and flakes \cite{sahu2}, it was found that the interlayer distance and the appearance of edge magnetism is sensitive to the particular local or semi-local approximation used. For the sake of consistency and meaningful comparisons, the same generalized gradient approximation (GGA) \cite{perdew} was used to capture the edge magnetism with the 
fixed interlayer distance of 0.335 nm. 

We note that van der Waal's interaction, which anchors the layers of graphene together, is not addressed within Kohn-Sham DFT and recently there have been several attempts to include these interactions seamlessly in the density functionals \cite{kronik,timo}, which could capture the interlayer distance between graphene layers accurate to within 5$\%$. 
It was shown \cite{timo}, however, that although van der Waals interaction contribute to interlayer distance dependent total energy, it has weak influence on overall band structure at a given distance, thus we neglect the van der Waals interaction in this study.

The unsaturated carbon $\sigma$-orbitals were passivated with hydrogen atoms and the C-H distance was fixed at 0.1084 nm. Therefore, our results are appropriate in situations where ribbons are cut in a hydrogen environment. 

In graphene sheets, there are three distinct stacking arrangements, labeled by A, B and C, as shown in Fig.~\ref{fig:fig1}(a).
For nanoribbons, we consider two types of edges (armchair and zigzag), each with two edge alignments namely $\alpha$- and $\beta$-alignments, as illustrated in Fig.~\ref{fig:fig1}(b). For more than two graphene layers, the stacking combinations multiply and each stacking sequence presents a 
particular 
band structure \cite{hongki1}. Density functional theory predicts that periodic AB stacking (Bernal) and periodic ABC stacking (rhombohedral) are almost energetically degenerate in bulk graphite \cite{gonze}. Recently, ABC stacking was realized on a SiC(0001) substrate \cite{norimatsu} experimentally, and electric field-induced gap opening was predicted theoretically in ABC-stacked trilayer graphene as well as ABCA-stacked tetralayer graphene \cite{fan,latil,aoki}. Thus, for multilayer ribbons with more than 3 layers, we only focus on periodic AB and ABC stacking sequences.

\begin{figure}[ht]
\includegraphics[width=1.0\linewidth]{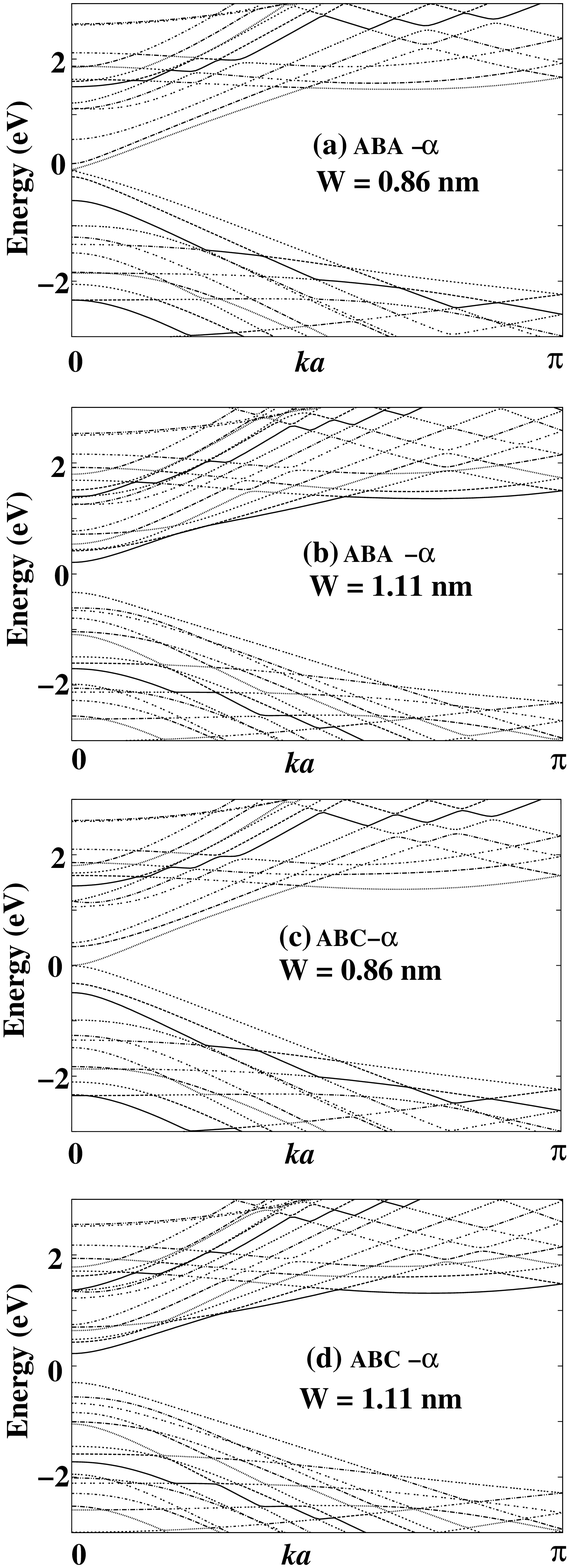}
\caption{ Energy band structure of trilayer armchair nanoribbons for (a) ABA-$\alpha$, metallic nanoribbon, (b) ABA-$\alpha$, semiconducting nanoribbon (c) ABC-$\alpha$, metallic nanoribbon, and (d) ABC-$\alpha$, semiconducting nanoribbon. The chosen widths for metallic and semiconducting ribbons are 0.86 nm and 1.11 nm which corresponds to $N$=8 and $N$=10 respectively, where $N$ is the number of carbon chains along the width direction.
}
\label{fig:fig2}
\end{figure}

\begin{figure}[ht!]
\includegraphics[width=0.65\linewidth]{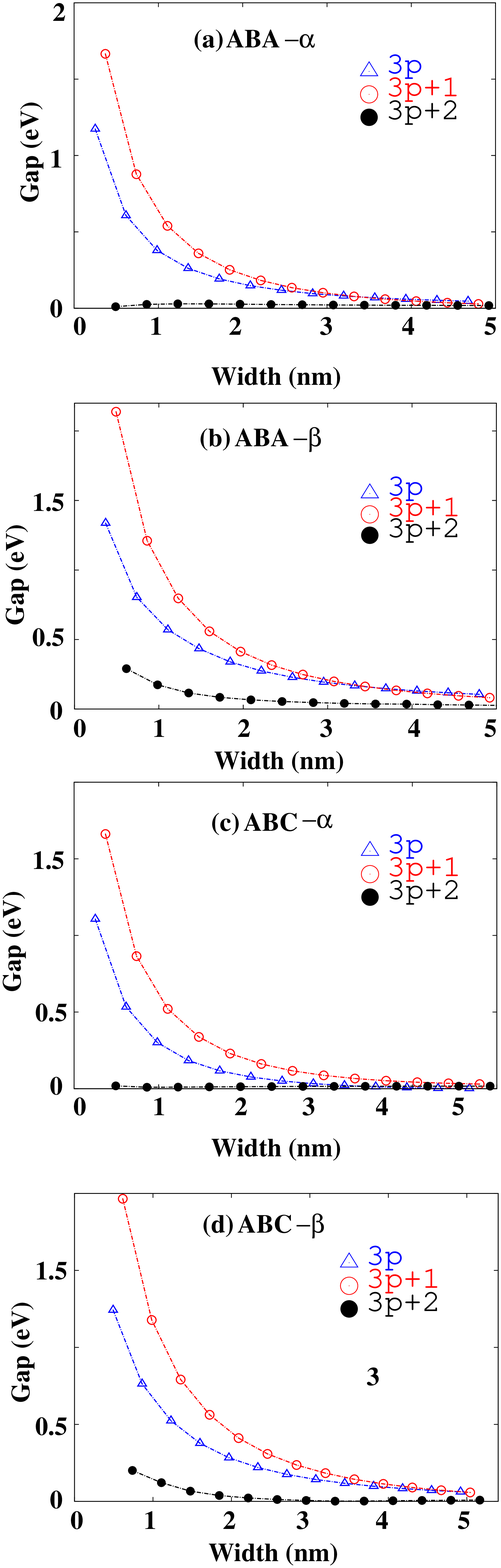}
\caption{ (Color online)  Variation of the energy gap with widths of (a) ABA-$\alpha$, (b) ABA-$\beta$, (c) ABC-$\alpha$, and (d) ABC-$\beta$-aligned nanoribbons. Three classes of the nanoribbons are clearly seen in (a)-(d) and  specified by $N=3p$, $3p+1$ and $3p+2$ where {\it N} is the number of carbon chains along the width direction. Here {\it p} =1,2,$\cdots$,13, which translate to nanoribbons with widths 
up to 5 nm.}
\label{fig:fig3}
\end{figure}

To establish the ground state magnetic order for the zigzag ribbons, we tested both narrow and wide ribbons with 
non-magnetic, ferromagnetic and antiferromagnetic order between the layers, while we set ferromagnetic coupling along each edge and antiferromagnetic coupling between the two edges within the same layer, as predicted theoretically \cite{fujita}. We find that interlayer antiferromagnetic order has lower energy than ferromagnetic, non-magnetic, or non-collinear magnetic order in the ribbons. We also find that the same ground state is reached with other forms of semi-local exchange-correlation potentials such as PBE \cite{burke}, PBESol \cite{perdew1}, RPBE \cite{zhang}. Therefore, for calculating band structures and other related quantities in this article, we consider interlayer antiferromagnetic order as a magnetic ground state of multilayer graphene ribbons. 
    
The nanoribbons were placed in a supercell with vacuum regions next to the width and the stacking direction. For trilayer ribbons, we used a 1.5 nm vacuum region and for tetralayer, pentalayer and hexalayer ribbons, 2 nm, 2.5 nm  and 3 nm vacuum regions were used, respectively. 
The necessity for taking different vacuum sizes for increasing number of layers was guided by the need to avoid intercell interaction in DFT in order to treat the multilayer stack as an isolated system and also the saw-tooth type implementation of electric potential in the DFT code we used. 

The bulk trilayer graphene calculations \cite{fan} suggested an external electric field, applied perpendicular to the ribbon layers, as high as 4 V/nm is necessary in order to achieve saturation of the gap. Therefore, we use a maximum external electric field strength of 5 V/nm to study its effect on the multilayer ribbon gaps. We used 68 {\bf k}-points in the irreducible part of the Brillouin Zone (BZ) 
and kinetic energy cut-off of 475 eV.
 The convergence of the calculations was tested with respect to a denser {\bf k}-point mesh, larger energy cut-offs as well as larger vacuum sizes. 

\section{Armchair graphene sheets} 
In this section, we discuss the electronic structure of multilayer armchair ribbons and the width dependence of energy gaps.

\subsection{Trilayer nanoribbons}

Figure \ref{fig:fig2} shows energy band structures of armchair ribbons stacked in ABA and ABC fashion. The Fermi energy is placed at zero. Like in monolayer and bilayer armchair nanoribbons, both metallic and semiconducting trilayer ribbons are possible. Note that for the metallic ABA nanoribbon (Fig.~\ref{fig:fig2}(a)) we find one linear and one quadratic band, while for metallic ABC nanoribbon (Fig.~\ref{fig:fig2}(c)) there is one cubic band near the Fermi energy. The appearance of the low energy states and corresponding energy dispersions are consistent with the energy spectrum analysis in bulk multilayer graphene \cite{hongki1}.

We now discuss the width dependence of the ribbon gaps. We consider 
the ribbons with maximum widths up to 5 nm.
Figure \ref{fig:fig3} shows the width dependent gap variation of trilayer armchair ribbons for both $\alpha$- and $\beta$-alignments in ABA and ABC stackings. 
Similarly as monolayer\cite{fujita} and bilayer ribbons\cite{sahu1},
three classes of ribbons are clearly seen in all cases,
two semiconducting ($N=3p, 3p+1$) and one metallic ($N=3p+2$), where $N$ denotes the number of carbon chains along the width direction and $p$ is an integer number. 
(Here, because of relatively smaller gap size, we call $N=3p+2$ ribbons as {\it metallic} though for small $p$, the ribbons are actually semi-conducting.)

Note that compared to bilayer armchair nanoribbons, these gaps are consistently smaller \cite{sahu1}. As the number of layers increases, more energy bands appear near the Fermi level due to the coupling between the layers at the DFT level, which reduces the gap size. 

\subsection{Multilayer nanoribbons}

\begin{figure}[ht]
\includegraphics[width=0.8\linewidth]{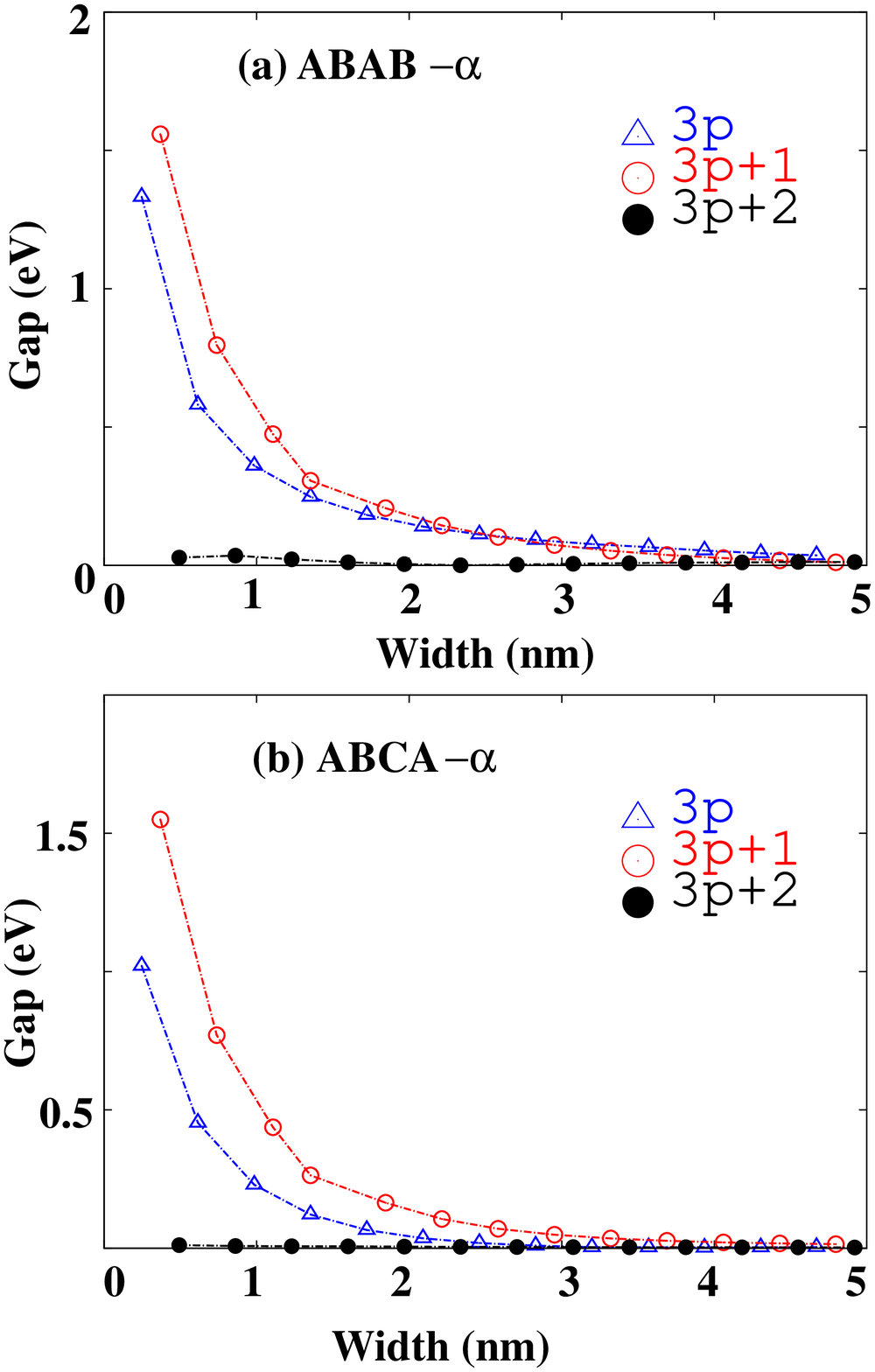}
\caption{ (Color online)  Variation of the energy gap with widths of tetralayer nanoribbons (a) ABAB-$\alpha$, (b) ABCA-$\alpha$. Three classes of 
nanoribbons are clearly seen. These classes are specified by $N=3p$, $3p+1$ and $3p+2$ where {\it N} is the number of carbon chains along the width direction. Here {\it p} =1,2,$\cdots$,13, which translate to nanoribbons with widths 
up
to 5 nm.}
\label{fig:fig4}
\end{figure}

For tetralayer graphene, we consider periodic AB and ABC stacking sequences
with widths up to 5 nm.
As with
mono-, bi- and tri-layer graphene ribbons, there exists three classes of ribbons in both stacking sequences, as shown in Fig.~\ref{fig:fig4}. Only ribbons with $\alpha$-alignment are considered here; the $\beta$-aligned ribbons show similar behavior (figures not shown). 

It is clearly seen from Fig.~\ref{fig:fig4} that the gaps are consistently becoming smaller for a given width, compared to bilayer \cite{sahu1} and trilayer ribbons (Fig. \ref{fig:fig3}). 

The origin of this decrease in gaps may again be attributed to appearance of more energy states, near the Fermi level, as discussed in the previous section. 
Although explicitly not shown, we believe that in thicker ribbons (with more than four layers), the width dependence of gap may show three classes of ribbons, albeit with gaps smaller than the gaps in ribbons with less than four layers.

\section{Zigzag graphene sheets}

In this section, we study zigzag edged multilayer ribbons and the induction of an energy gap due to edge magnetism. We first describe the 
interesting  
electronic structure and edge magnetism in trilayer zigzag nanoribbons, and then extend our discussions to multilayer ribbons.

\subsection{Trilayer nanoribbons}

\begin{figure}[ht]
\includegraphics[width=1.0\linewidth]{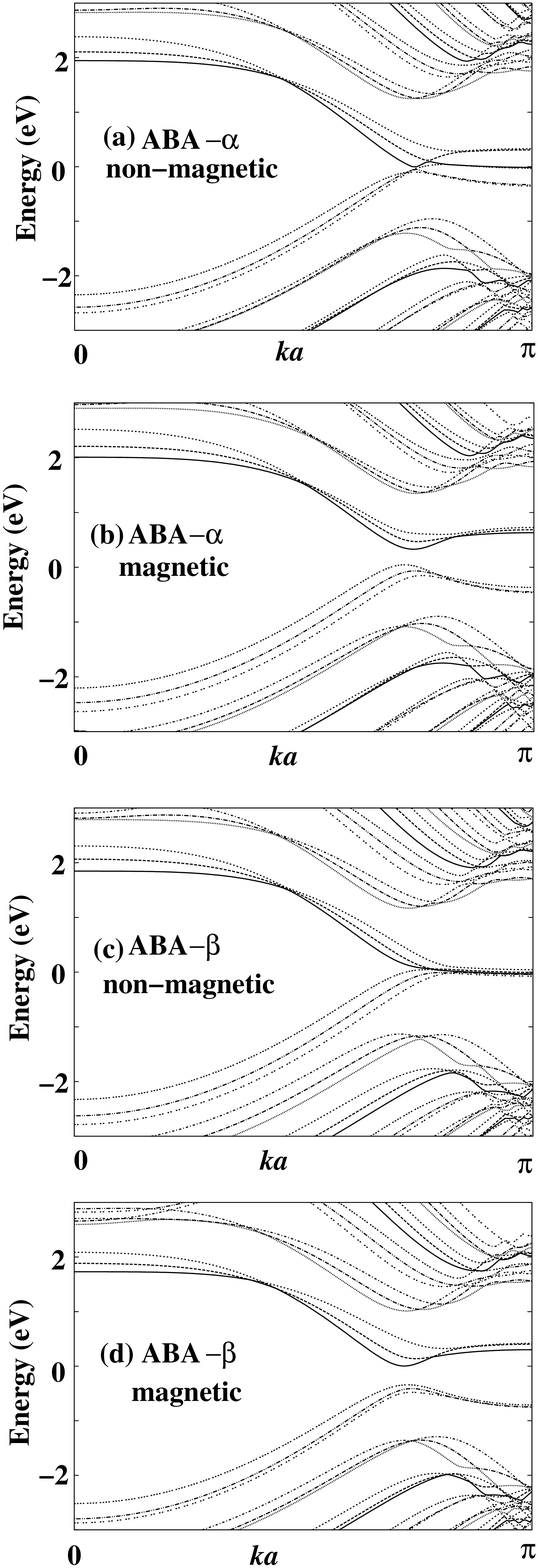}
\caption{ Energy band structure of ABA-stacked trilayer zigzag nanoribbons with $\alpha$ and $\beta$ alignments.
Panels (a) and (b) are, respectively, for non-magnetic and magnetic $\alpha$-aligned ribbons, and panels (c) and (d) are for $\beta$-aligned ribbons. The width is 
1.7 nm which corresponds to $N$=8 where $N$ is the number of zigzag chains along the width directions.
}
\label{fig:fig5}
\end{figure}

Zigzag ribbons, due to their edge states, show magnetic order and their ground states are predicted to have the interlayer antiferromagnetic order, as discussed in Sec. II. Figure \ref{fig:fig5} shows the non-magnetic and magnetic band structure of a 1.7 nm wide trilayer zigzag ribbon stacked in ABA-fashion with both $\alpha$- and $\beta$-alignments. We find 
similar band structures for ABC-stacked ribbons. In non-magnetic ribbons with $\alpha$-alignment (Fig. \ref{fig:fig5}(a)), a flat band appears at the Fermi level and as a result, the system is unstable
and becomes magnetic
due to the large density of states. Several other flat bands also appear away from the Fermi level. Edge magnetic order induces an energy gap by breaking the flat band degeneracy (Fig. \ref{fig:fig5}(b)).

In non-magnetic $\beta$-aligned ribbons (Fig. \ref{fig:fig5}(c)), several flat bands appear together at the Fermi level and  as a result, the system is again unstable
and becomes magnetic
due to the presence of comparatively 
large density of states. When we take into account the edge magnetic order, 
an energy gap is opened at the Fermi level.
(Fig. \ref{fig:fig5}(d)). The flat bands and magnetically induced energy gap are also seen in bilayer graphene nanoribbons (see Fig. 5 and 6 in Ref. 18). The number of flat bands at the Fermi level or away from it depends on the number of graphene layers.

\begin{figure}
\includegraphics[width=0.8\linewidth]{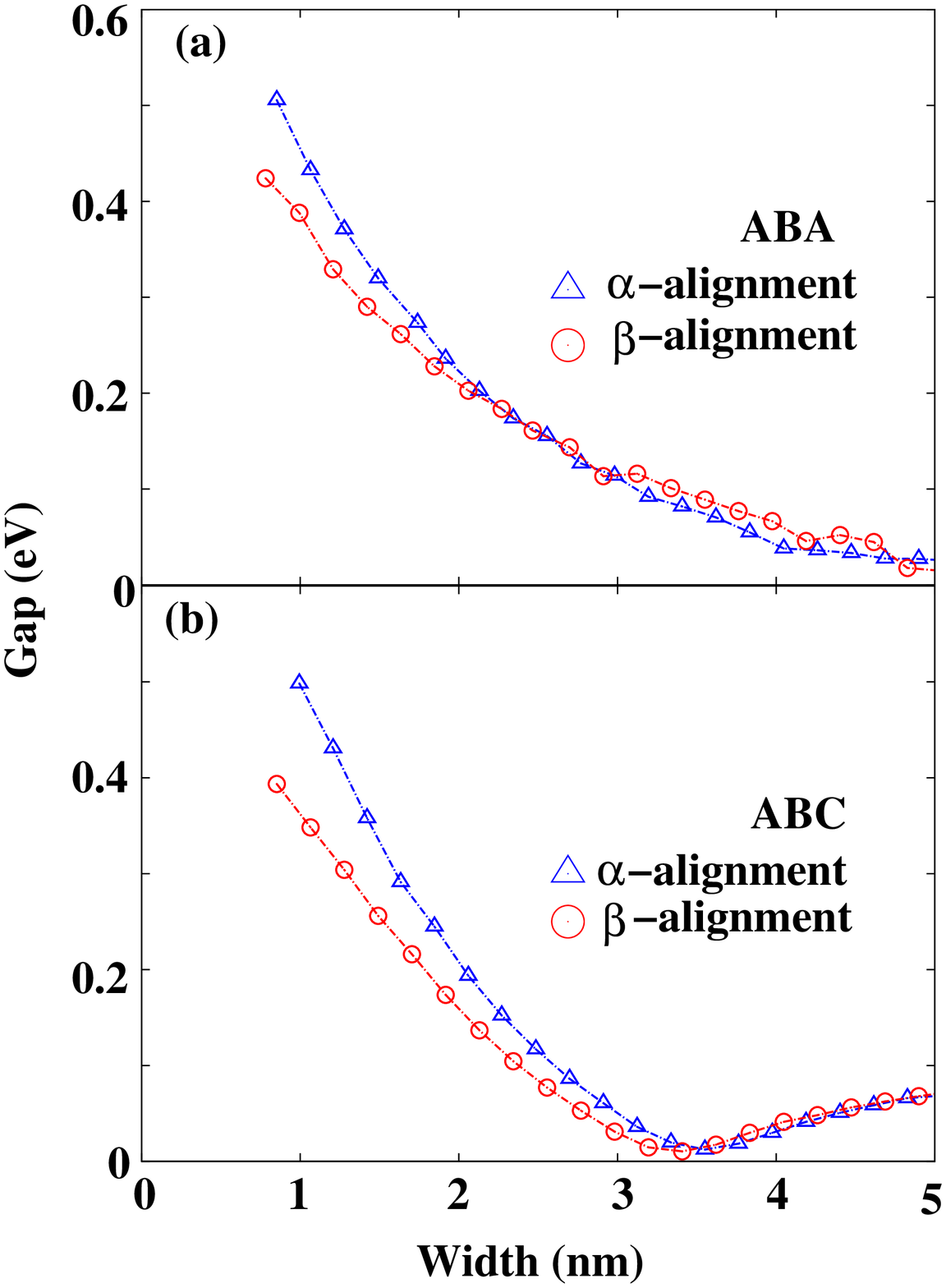}
\caption{Variation of energy gap with the widths of (a) ABA and (b) ABC-stacked trilayer zigzag nanoribbons.
}
\label{fig:fig6}
\end{figure}

We now discuss the variation of the magnetically induced gaps with the ribbon width. We consider a maximum width of 5 nm. Figures \ref{fig:fig6}(a) and (b) show a monotonic decrease of the gap with the ribbon width in both ABA- and ABC-stacked ribbons, respectively, each with $\alpha$- and $\beta$-alignments. Due to magnetism, for comparable widths, 
the energy gaps of zigzag ribbons are larger than those of corresponding armchair ribbons. 
For example, for an ABA-stacked ribbon with a width of 1.2 nm (which translates to $N$=5 and $N$=11 carbon chains for the zigzag ribbon and the armchair ribbon, respectively) and $\alpha$-alignment, the gap is 0.433 eV (0.118 eV) for the zigzag (armchair) ribbon. The increase of the gap with increasing width after 3 nm in ABC-stacked ribbons is somewhat surprising. We increased the vacuum region from 1.5 nm to 3 nm and 5 nm, and repeated the calculations for these widths but again find the same increasing trend.

\subsection{Multilayer nanoribbons}

We now extend our discussions to non-magnetic and magnetic multilayer ribbons, which show distinct features in their band structures depending on the edge alignment and whether the number of layers is odd or even. For illustrative purposes, we consider only AB-periodic ribbons with $\alpha$-alignment. $\alpha$-aligned ABC-periodic ribbons show similar behavior. For $\beta$-aligned ribbons, both periodic stacking sequences show several flat bands appearing together at the Fermi level as in trilayer counterparts, and do not show any qualitative change with the number of layers.

\begin{figure}[ht]
\includegraphics[width=1\linewidth]{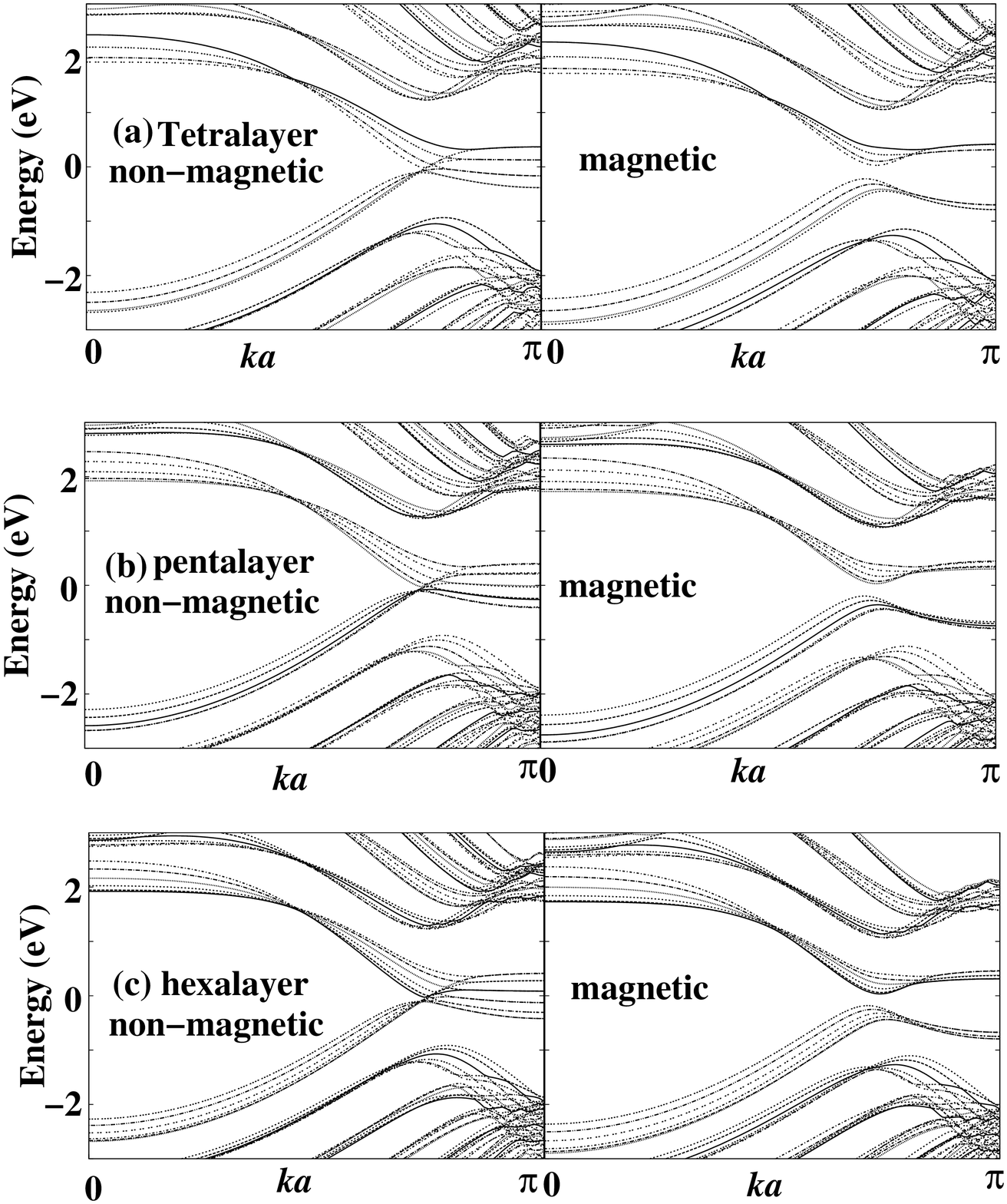}
\caption{Energy band structure of multilayer zigzag nanoribbons for ABA-stacked (a) tetralayer (b) pentalayer and (c) hexalayer with $\alpha$-alignment. The left panels show the nanoribbons with non-magnetic edge atoms and the right panels 
show the nanoribbons
with edge atoms ordered antiferromagnetically layerwise. The width is chosen as 1.7 nm.
}
\label{fig:fig7}
\end{figure}

Figure \ref{fig:fig7} shows non-magnetic (left panels) and magnetic (right panels) band structures of tetralayer, pentalayer and hexalayer $\alpha$-aligned ribbons in ABA-stacking sequence. In non-magnetic ribbons with an even number of layers (such as in tetralayer and hexalayer ribbons), several flat bands appear away from the Fermi level, whereas for an odd number of layers (such as in pentalayer ribbons), one degenerate flat band appears at the Fermi level with several other flat bands away from it. The number of such flat bands depends on the number of layers considered. The appearance of flat bands at the Fermi level results in magnetic instability and as a result magnetic order induces gaps in all cases (Fig. \ref{fig:fig7}, right panels). 

With the layer antiferromagnetic order, we expected to find magnetically induced gaps larger in $\alpha$-aligned odd-layered zigzag ribbons compared to the $\alpha$-aligned even-layered ribbons due to a flat band appearing directly at the Fermi level.
We also expected that $\beta$-aligned ribbons exhibit larger gaps compared to $\alpha$-aligned ribbons due to several flat bands at the Fermi level. 
But we did not find any such trends. 
For comparison, we tested ribbons with layer \textit{ferromagnetic} order (both $\alpha$- and $\beta$-alignments), 
and found that in $\beta$-aligned ribbons, the magnetically induced gaps are consistently larger than their $\alpha$-counterparts, albeit 
still
no trend in gaps of odd-even layers is seen either in this case.     

\begin{figure}
\includegraphics[width=0.8\linewidth]{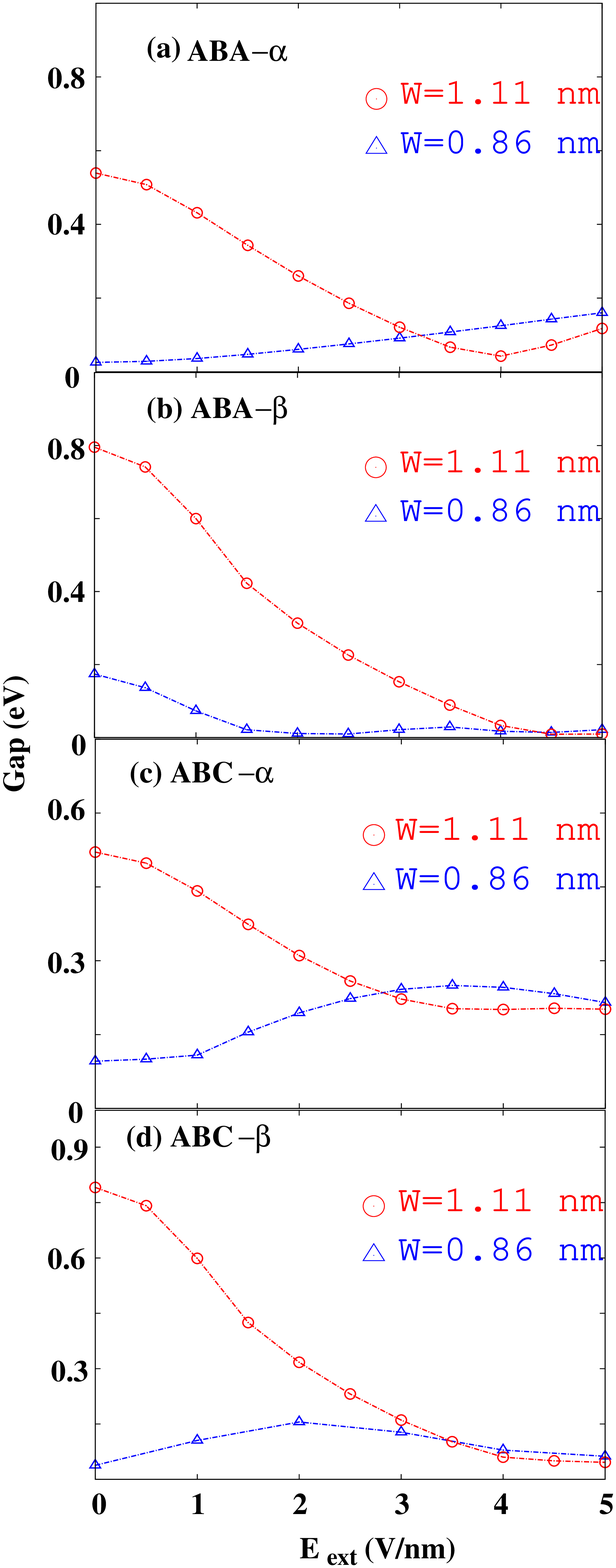}
\caption{ (Color online) Variation of the energy gap with perpendicular external electric field for ABA-stacked and ABC-stacked trilayer armchair ribbons. Panels (a) and (b) are for $\alpha$- and $\beta$-aligned ABA-stacked ribbon respectively whereas panels (c) and (d) are, respectively for ABC-stacked $\alpha$- and $\beta$-aligned ribbons. Both metallic (W = 0.86 nm) and semiconducting (W=1.11 nm) nanoribbons are considered which corresponds to $N$=8 and $N$=10 respectively, where $N$ is the number of armchair carbon chains along the width direction.}
\label{fig:fig8}
\end{figure}

\section{Electric field effects on the gaps} 

This section deals with the effect of external electric fields, applied perpendicular to the ribbons, on the confinement and edge magnetism induced gaps \cite{sahu3}. We choose representative metallic and semiconducting ribbons to illustrate the effect. Figure \ref{fig:fig8} shows the variation of energy gap with external electric fields in ABA-stacked armchair 
ribbons
with $\alpha$- and $\beta$-alignments (a-b) respectively, and corresponding alignments in ABC-stacked ribbons (c-d). We consider a maximum field strength of 5 V/nm. A gap is induced by electric fields in ABC-stacked metallic ribbons, increasing with small fields and then showing signs of saturation. This is consistent with the prediction of a gap opening in ABC-stacked bulk trilayer graphene and the electric field effects on the induced gaps \cite{fan}. In ABA-stacked metallic ribbons, gaps remain closed or constant for all electric field strengths. In all cases, for semiconducting ribbons initially with a large gap, we find monotonic decrease of gaps with increasing electric field. This behavior is similar to electric field effects in bilayer armchair ribbons \cite{sahu1}.

Our calculations on bilayer nanoribbons suggested a critical band gap of 0.2 eV above and below which the external electric field change the sign of the gap values. Although such critical gap value is not explicitly proved here, we believe that a critical gap may also exist in multilayer nanoribbon stacks. Therefore, we expect that these gaps (above the critical gap) to decrease with increasing strength of external electric fields applied perpendicular to the layers. 

\begin{figure}
\includegraphics[width=0.8\linewidth]{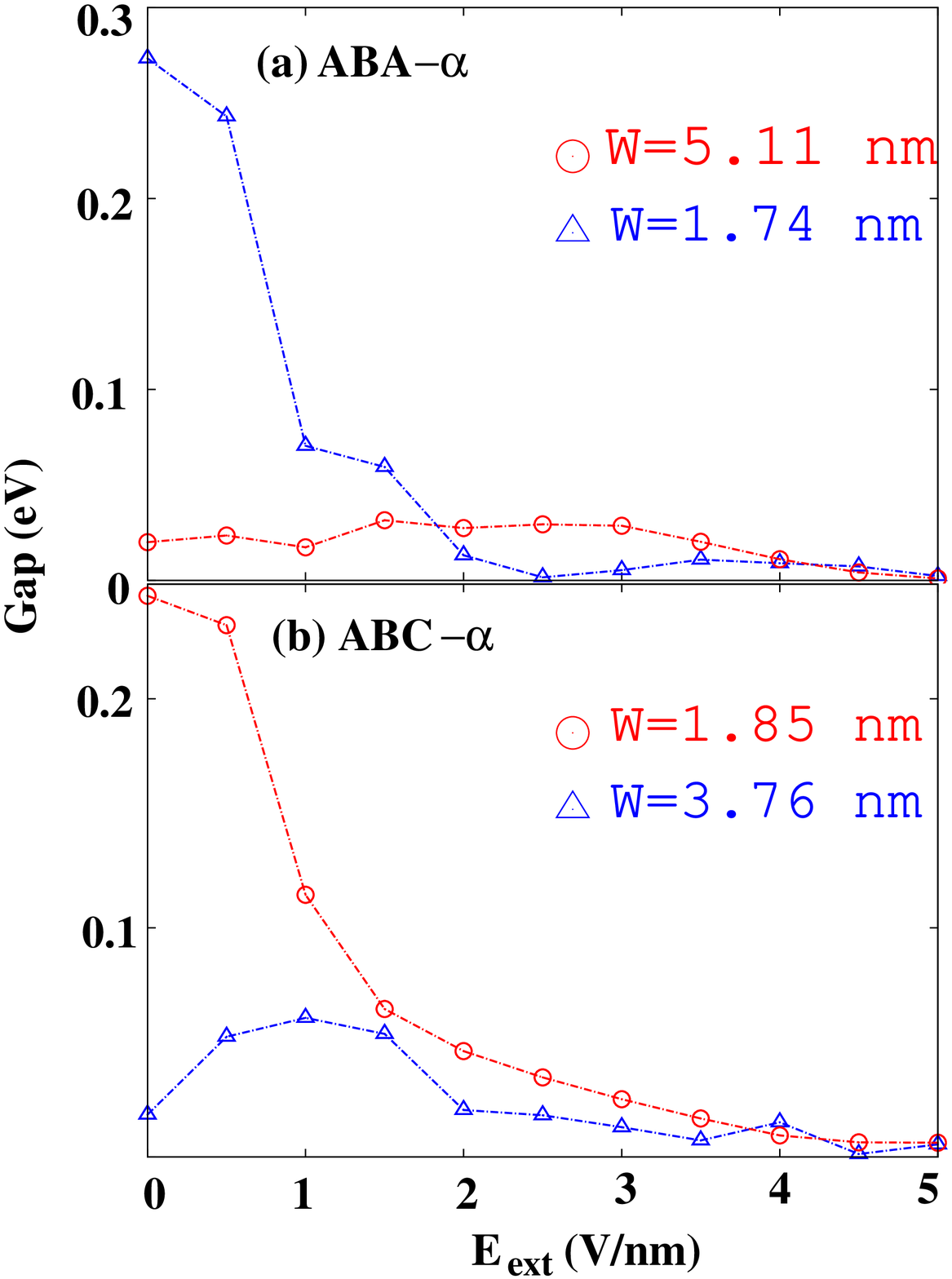}
\caption{ (Color online)
Variation of the energy gap with perpendicular external electric field for
(a) ABA and (b) ABC-stacked trilayer zigzag nanoribbons with $\alpha$-alignments. $\beta$-aligned nanoribbons show similar behavior. Both narrow (W = 1.74 nm and 1.85 nm) and wide  (W= 5.11 nm and 3.75 nm) nanoribbons are considered.}
\label{fig:fig9}
\end{figure}

To study the electric field effects in zigzag ribbons, we choose both wide and narrow gap ribbons. Only $\alpha$-aligned ribbons in both stacking sequences are considered. $\beta$-aligned ribbons show similar behavior (figures not shown). Figure \ref{fig:fig9} shows the variation of energy gap of the zigzag ribbon with respect to electric field strengths (chosen up to a maximum of 5 V/nm). In wide width ribbons which have initially a small gap, we find gaps increasing with increasing electric field strengths for ABC stacking while for ABA stacking, gaps remain closed or constant. In narrow width ribbons which have initially a large gap, 
the gap decreases with field.  
This trend is similar to that observed in bilayer zigzag nanoribbons \cite{sahu1}.

\section{Summary and Conclusions}
 
In summary, we have studied the electronic properties of armchair and zigzag multilayer graphene nanoribbons both with and without external electric fields using 
\textit{first principles} density functional-based electronic structure method. We consider both AB-periodic and ABC-periodic nanoribbons with two different edge alignments, referred to as $\alpha$- and $\beta$-alignments. Armchair edged multilayer ribbons exhibit three classes, two semiconducting and one metallic. The energy dispersion of metallic armchair nanoribbons near the Fermi energy is consistent with the chiral decomposition in bulk multilayer graphene. The gap in multilayer armchair ribbons, for a particular width, is found to be smaller than in the corresponding bilayer ribbons. We find that magnetically induced gaps in multilayer zigzag nanoribbons and the modulation of the gap values depending upon the type of edge alignments. ABC-periodic nanoribbons with a small gap show enhancement in the gap values with increasing electric field strengths, while AB-periodic nanoribbons with a small gap show no increase in the gap values with increasing electric fields. For nanoribbons with a large gap, the gap values decrease with increasing electric fields. 
In view of encouraging advances in the fabrication, control of graphene edges and observation of magnetic edge states, we believe that our studies of magnetism and electric field effects will be important for designing graphene based nanodevices.   

\acknowledgments

The authors acknowledge financial support from NRI-SWAN center.
The work done by H. M. has been supported in part by the NIST-CNST/UMD-NanoCenter Cooperative Agreement.
 B. S. acknowledges the allocation of computing time on NSF Teragrid machine {\it Ranger} (TG-DMR080016N) at Texas Advanced Computing Center and DOE leadership computing machine {\it Jaguar} at Oak Ridge National Laboratory (NEL003). 
The authors thank J. J. McClelland, P. Haney and K. Gilmore for their valuable comments.


\begin{thebibliography}{999}
\bibitem{sahu3} The applied electric field does not change the layer separation and the edge structures. Since we use GGA and fixed the interlayer distance at 
0.335 nm
for our calculations, we tested this hypothesis using LDA by relaxing all the atoms in the presence of maximum electric field strength of 5 V/nm we used for electric field effect studies.
\end{thebibliography}
\end{document}